\documentclass[10pt,journal,compsoc]{IEEEtran}
\usepackage{cite}
\usepackage{amsmath,amssymb,amsfonts}
\usepackage{algorithmic}
\usepackage{graphicx}
\usepackage{textcomp}
\usepackage{comment}
\usepackage{threeparttable}
\usepackage{hyperref}
\usepackage{threeparttable}
\usepackage{multirow}
\usepackage{booktabs}
\usepackage{upgreek}
\usepackage{algorithm}
\def\BibTeX{{\rm B\kern-.05em{\sc i\kern-.025em b}\kern-.08em
    T\kern-.1667em\lower.7ex\hbox{E}\kern-.125emX}}

\begin{document}

\title{Generalizable framework of eating episode detection on free-living wrist-worn wearable data}
\author{
	Chunzhuo Wang, Emma De Schuyteneer, Huaidong Du, Aiden Doherty, Elske Vrieze, and Bart Vanrumste
	\thanks{This work was supported in part by the Horizon Europe Research and Innovation Programme under Grant No. 101083388, the Flanders AI Research Programme, the Leuven.AI Institute, and the Research Foundation Flanders (FWO) under Travel Grant V451325N. The work of Emma De Schuyteneer was supported through a non-commercial academic research collaboration with the Interuniversity Microelectronics Centre (IMEC). The work of Aiden Doherty was supported in part by the Wellcome Trust (223100/Z/21/Z, 227093/Z/23/Z), the EPSRC Centre for Doctoral Training in Health Data Science (EP/S02428X/1), the British Heart Foundation Centre of Research Excellence (RE/18/3/34214), Cancer Research UK, the NIH Oxford–Cambridge Scholars Programme, Google, GSK, Boehringer Ingelheim, and the Danish National Research Foundation (Pioneer Centre for SMARTbiomed).}
	\thanks{Chunzhuo Wang, and Bart Vanrumste are with the e-Media Research Lab, and also with the ESAT-STADIUS Division, KU Leuven, 3000 Leuven, Belgium (e-mail: chunzhuo.wang@kuleuven.be; bart.vanrumste@kuleuven.be).}
	\thanks{Emma De Schuyteneer, and Elske Vrieze are with the Mind‐body Research, Biomedical Sciences Group, and also with the Leuven Brain Institute, KU Leuven, 3000 Leuven, Belgium (e-mail: emma.deschuyteneer@kuleuven.be;elske.vrieze@upckuleuven.be).}
	\thanks{Huaidong Du, and Aiden Doherty are with Nuffield Department of Population Health, University of Oxford, Oxford, UK, and also with Big Data Institute, Li Ka Shing Centre for Health Information and Discovery, University of Oxford, Oxford, UK (e-mail: huaidong.du@ndph.ox.ac.uk;aiden.doherty@ndph.ox.ac.uk).}
}

\maketitle
\begin{abstract}
Accurate assessment of eating behavior is essential for understanding and managing conditions such as eating disorders, obesity, and diabetes. Wearable-based food intake detection has shown considerable promise; however, most existing approaches are trained and evaluated using internal validation on a single dataset with fixed sensor orientation and known wearing hand, which substantially limits their generalizability to real-world settings. Furthermore, many existing approaches rely on both accelerometer (acc) and gyroscope (gyro) signals to achieve strong performance. However, gyro measurements may be unavailable in some real-world deployments due to battery constraints, and performance often degrades when only acc data are used. To address these limitations, we propose a generalizable framework for orientation- and wrist-invariant eating episode detection, with an auxiliary acc2gyro module to improve performance in acc-only settings. The framework is trained using fine-grained wrist-worn datasets and externally validated across three heterogeneous datasets: the Clemson All-Day (CAD) and Capture-24 datasets, as well as Physio-ED, a dataset collected from individuals with eating disorders. Across external evaluations, the proposed framework demonstrates robust performance despite substantial variations in sensor modality, wearing hand, participant population, and annotation protocols. Specifically, the framework achieved eating episode detection F1-scores of 0.751, 0.592, and 0.793 on CAD, Capture-24, and Physio-ED, respectively, with CAD performance exceeding recent state-of-the-art methods evaluated using internal validation only. Notably, this study provides the first external validation of eating episode detection in an eating disorder population. Additionally, the acc2gyro module improves the performance in acc-only settings. These findings demonstrate the potential of orientation- and wrist-invariant wearable sensing for scalable and clinically applicable assessment of eating behavior.
\end{abstract}


\section{Introduction}
\IEEEPARstart{D}{ietary} monitoring is critical for understanding food intake behavior, both in the general population and in specific clinical groups such as individuals with obesity, diabetes, or eating disorders \cite{b1,zandian2007cause}. Traditional dietary assessment tools, including 24-hour recall (24HR), food diaries, food frequency questionnaires (FFQ), and ecological momentary assessment (EMA) \cite{smyth2001use,allmeta2026same}, rely heavily on manual reporting. These methods are subjective, time-consuming, and prone to error, motivating the development of automated approaches that enable continuous, and unobtrusive dietary monitoring without human annotation in daily life.

Recent advances in artificial intelligence and wearable sensing have led to a variety of approaches for automated food intake monitoring \cite{wang2025scoping, b20, b17, m2,s18}. In particular, wrist-worn inertial measurement units (IMUs) have shown promise for detecting eating gestures and episodes \cite{m1,b51, wang2024detection,tang2023detecting}. However, most existing methods to detect eating episodes using wrist-worn IMUs are trained and evaluated on a single dataset \cite{10584254}, or rely on pretraining on one dataset followed by fine-tuning on another \cite{tang2023detecting}. Consequently, it remains unclear whether these models generalize to new populations or devices, and external validation is often lacking. This limitation restricts the applicability of these methods in domains such as precision nutrition, public health, and eating disorder research.

Evaluating automated food intake detection methods in external datasets is critical for real-world, large-scale monitoring, but presents significant challenges. Models trained under controlled conditions typically require a specific sensor type, fixed orientation, and knowledge of which wrist the sensor is worn on. Variations in sensor placement and wearing hand can substantially degrade performance, making it difficult to deploy existing models in broader application scenarios.

Additionally, while many existing intake activity detection studies rely on 6-axis IMU sensors, comprising 3-axis accelerometers (acc) and 3-axis gyroscope (gyro), long-term real-world monitoring scenarios often employ only 3-axis accelerometers due to battery constraints \cite{mpu9250_datasheet}. The absence of gyroscope data removes critical rotational information, leading to a substantial degradation in model performance.

To address these challenges, we propose a generalizable framework for eating episode detection that operates across datasets with unknown sensor orientation, unknown wearing hand, and heterogeneous participant populations. The main contributions of this study include:
\begin{itemize}
	\item We developed a multi-branch framework that enables robust eating gesture recognition under unknown sensor orientation and wearing hand conditions without requiring user-specific calibration. The framework further treats acc and gyro signals as independent input streams to improve modality-specific feature extraction.
	\item We introduced an acc2gyro virtual sensing module that reconstructs gyroscope-like representations from acc signals, reducing performance degradation in acc-only settings where gyroscope measurements are unavailable due to energy or hardware constraints.
	\item We performed extensive external validation across heterogeneous datasets, including the Clemson All-Day (CAD) dataset from the US \cite{10.1145/3407623}, the Capture-24 dataset from the UK \cite{chan2024capture}, and Physio-ED from Belgium \cite{10.3389/fnins.2019.00606}. The proposed framework surpassed previously reported state-of-the-art performance on CAD despite prior methods relying on internal validation only. Additionally, to our knowledge, this study provides the first external validation of wrist-worn eating episode detection in an eating-disorder population.
\end{itemize}

\section{Related Work}
\subsection{Eating gesture and episode detection}

Eating gesture detection has been extensively explored using a variety of sensors. An eating gesture is defined as a sequence of continuous movements, from using the hand to pick up food to moving the hand away from the mouth. Vision-based systems have been applied to identify bite events \cite{m3,lb7} as well as food types \cite{b20}. Acoustic sensing has also been used to capture chewing sounds \cite{b7,tan2026nutriear}. Mertes et al. \cite{b12} introduced a smart plate embedded with strain gauges that estimates bite events by tracking changes in the food weight. In our earlier Eat-Radar study \cite{m16}, we validated a radar-based framework for eating detection. In addition, physiological sensing techniques such as photoplethysmography (PPG) \cite{s49a} and electromyography (EMG) \cite{b17} have been investigated for bite detection. Among these sensors, wrist-worn IMU sensors have become particularly popular due to their minimal user burden and high user acceptability. Dong et al. \cite{b28} proposed a rule-based method that detects bite events based on wrist rotational velocity. Shen et al. \cite{m18} subsequently evaluated this method using the Clemson dataset. Kyritsis et al. \cite{m1} introduced an end-to-end deep learning framework combining convolutional neural networks and long short-term memory networks (CNN-LSTM) for bite detection on the FIC dataset. Building upon this line of work, Rouast et al. \cite{m13} proposed a ResNet-LSTM architecture and evaluated it on the OREBA dataset. Wei et al. \cite{s28} later developed an energy-efficient solution by integrating an optimized multicenter classifier (O-MCC) to recognize eating gestures while maintaining low inference latency. Although these studies have demonstrated encouraging results, most of them primarily focus on detecting eating gestures within relatively short meal sessions.

Building on eating gesture detection during time-limited meal sessions, recent research has explored a more challenging setting: detecting eating gestures in (near-)free-living environments lasting $\geq{6}$ h to identify eating episodes. An eating episode is defined as a continuous segment of time during which repeated eating gestures occur, beginning with the first detected gesture and ending when no further eating gesture is observed. Bedri et al. \cite{s18} introduced the FitByte, an eyeglass-based sensing platform designed to capture eating gestures. In their approach, individual eating gestures occurring within a 5-minute interval were merged to form an eating episode. Zhang et al. \cite{b25} proposed the Necksense, which detects chewing sequences and groups them into eating episodes using the density-based spatial clustering of applications with noise (DBSCAN) algorithm \cite{s42}. Kyritsis et al. \cite{m1} detected eating gestures on the FreeFIC dataset and applied a Gaussian filter to the bite sequence to derive meal regions. These methods follow a 'bottom-up' paradigm, in which low-level eating events such as chewing, swallowing, or hand-to-mouth gestures, are first detected and subsequently aggregated into higher-level eating episodes using heuristic or clustering-based strategies.

In contrast, another line of work adopts a top-down approach \cite{sharma2022top}, aiming to infer eating episodes directly from raw sensor signals. A typical pipeline segments full-day sensor recordings into minute-level windows and applies machine learning models to classify whether each segment corresponds to an eating episode. For example, Sharma et al. \cite{10.1145/3407623} collected a large-scale dataset consisting of 354 days of wrist-worn IMU recordings from 351 participants. Doulah et al. \cite{m2} proposed the AIM-2 system, which uses eyeglasses equipped with a camera and a 3-axis accelerometer to detect eating episodes.

Together, these studies demonstrate the feasibility of detecting eating gestures and episodes using wearable sensors, laying the foundation for automated eating behavior monitoring.

\subsection{Assumptions on sensor placement}
Despite the demonstrated success, most existing approaches implicitly rely on strong assumptions regarding sensor placement. In particular, data are typically collected with a fixed and predefined sensor orientation, and the wearing hand (i.e., dominant or non-dominant wrist) is known \cite{m1,s28}. Under this setup, models are trained and evaluated on signals whose coordinate frames and motion patterns are aligned, simplifying both feature extraction and classification. This assumption-driven pipeline has been adopted in our previous work on the Full-Day I (FD-I) dataset\cite{10584254}, where participants wore Shimmer3 IMUs on both wrists with fixed sensor orientations, as well as in studies using the meal-only OREBA dataset \cite{m13} and the work on the CAD dataset \cite{10.1145/3407623}.

However, such assumptions are difficult to guarantee in unconstrained, real-world settings, where users may wear devices on either wrist, or attach sensors with varying orientations due to differences in device design or user behavior. These variations introduce changes in inertial signals, even for identical eating motions, leading to degraded model performance when methods trained under controlled configurations are applied to data collected under different conditions.

\subsection{Missing gyroscope in real-world monitoring}
Wrist-worn eating gesture and episode detection approaches commonly leverage inertial sensing, where many studies combine acc and gyro signals to capture both linear and rotational wrist motion \cite{m1,m13,s28}. While acc-only approaches have also been explored in earlier stage \cite{thomaz2015practical}, recent methods more frequently adopt multi-sensor (acc+gyro) configurations to improve motion representation. In contrast, large-scale and long-term real-world deployments, such as population studies (e.g., UK Biobank \cite{yuan2024self}, US NHANES \cite{shim2023wearable}, and China Kadoorie \cite{chen2023device}) and widely used wearable platforms (e.g., Empatica, Axivity, Fitbit), often rely on acc-only sensing due to energy and scalability constraints. This discrepancy between commonly used multi-sensor research settings and acc-only deployment scenarios raises the question of whether methods developed under richer sensing conditions can generalize to real-world long-term monitoring settings.

\subsection{Cross-dataset generalization and external validation}
Cross-dataset generalization has received increasing attention in wearable-based human activity recognition (HAR) \cite{hong2024crosshar, cai2025towards}, yet remains relatively underexplored in the context of eating episode detection. Most existing studies evaluate model performance using subject-independent splits within a single dataset, where sensor type, wearing configuration, and annotation protocols are consistent across training and testing sets \cite{10584254,10.1145/3407623,m1}. Such evaluations assess within-dataset performance and provide limited insight into model robustness under heterogeneous data collection conditions.

A limited number of studies have explored cross-dataset transfer for eating behavior recognition by pretraining models on one dataset and subsequently fine-tuning or adapting them to another. Tang et al. \cite{tang2023detecting} trained an eating gesture classifier using multiple in-meal datasets and then retrained and evaluated the model on the free-living CAD dataset. While such approaches demonstrate the feasibility of knowledge transfer, they typically rely on partial retraining and assume access to labels from the target domain. Another underlying limitation is the lack of a standardized definition and annotation protocol for eating gestures across datasets, leading to inconsistencies in label semantics. Consequently, models trained on one dataset often cannot be directly transferred without retraining on the target dataset. Therefore, their applicability to fully independent external datasets remains limited.

These limitations underscore the need for a deployable and generalizable eating episode detection framework that can operate on independent datasets collected under diverse real-world configurations, including variations in sensor placement, wearing hand, and participant population characteristics.

\section{Datasets Description and Preprocessing}

\subsection{Full-day datasets}
\subsubsection{Full-Day-I (FD-I)}
The public available FD-I dataset introduced in our earlier work \cite{10584254} consists of IMU data collected over 34 days from 34 participants (24 males/10 females, age range: 18-35 years) in Leuven, Belgium. During data collection, participants were encouraged to carry out their daily routines while wearing two Shimmer3 IMU wristbands, one on each wrist. Each sensor integrates a 3-axis accelerometer and a 3-axis gyroscope, both sampled at 64 Hz. To obtain ground-truth annotations, a research assistant accompanied the participants and used a camera to document their activities, enabling manual labeling of all eating and drinking gestures. In total, the dataset includes 252 h of two-hand IMU recordings, covering 74 eating episodes with 4,568 eating gestures. The dataset captures four common eating styles, fork and knife, chopsticks, spoon, and hand-based eating. The recorded eating scenarios span multiple real-world environments, including participants’ homes, restaurants, and work-related spaces such as libraries, university learning centers, and campus rest areas.

\subsubsection{Clemson-All-Day (CAD)}
The CAD dataset is currently the largest publicly available dataset of wrist motion recordings for eating behavior analysis \cite{10.1145/3407623}. It comprises data from 351 participants (137 males/214 females; age: $28 \pm 12$ years) recruited from the surrounding areas of Clemson, USA. Each participant wore a Shimmer3 device on the wrist of their dominant hand for an entire day. The device captured 3-axis accelerometer and gyroscope signals at a sampling rate of 15 Hz, producing 4,680 h of sensor data. Participants were not constrained in terms of food choices, utensils, or eating habits during the recording process. To provide ground-truth labels, participants were instructed to press a button on the device at the start and end of each meal or snack.  In total, 1,107 distinct eating episodes were identified.

\subsubsection{Capture-24}
The Capture-24 dataset \cite{chan2024capture} comprises 2,562 h of wrist-worn accelerometer recordings collected from 151 participants in Oxfordshire, United Kingdom (52 males/99 females, age range: 18–53+ years), with each participant monitored for approximately 24h. Data were acquired using the Axivity AX3 accelerometer with a sampling rate of 100 Hz. During the study, participants also wore a camera around the neck, which automatically captured images every 20–40 seconds. Although the time-stamped images from the wearable camera are not publicly released, they were used by researchers to annotate the accelerometer signals during participants’ waking periods, and these annotated accelerometer data are publicly available. For this dataset, the wristband was worn on the dominant hand, but information regarding whether the dominant hand was left or right was unavailable. Further, the orientation of the sensor on the hand is unknown.
\subsubsection{Physio-ED}
The Physio-ED dataset \cite{10.3389/fnins.2019.00606} contains accelerometer data from both hands of 57 female participants with eating disorders (age: $27 \pm 8$ years) receiving treatment for their eating disorder at a psychiatric center in Belgium. The study was approved by the Ethics Committee of UZ Leuven (approval number S59491), and informed consent was obtained from all participants. In the treatment context, both inpatients and day-program patients wear the watch during the day while participating in therapy sessions, having meals together, and engaging in free time. Each participant is asked to wear a Chillband (IMEC vzw, Belgium) on both hands from 9 a.m. to 5 p.m., with a sampling frequency of 32Hz. Mealtimes are fixed in both the inpatient and day program settings.

\begin{table*}[t]
	\caption{Information of wrist-worn motion sensor datasets used for eating episode detection}
	\label{dataset_info}
	\begin{center}
		\scalebox{1}{
			\begin{threeparttable} 
				\begin{tabular}{l|l|l|l|l|p{1.7cm}|p{2cm}|l|l|l}
					\toprule
					\hline
					Dataset & \#Subjects & \#Hours & \#Gestures& \#Episodes& Wrist-worn info & Sensor info & Orientation &Label  &Environment \\ 
					\midrule
					FD-I\cite{10584254}&  34 & 252&4568& 74& two-hand & Shimmers3, 64Hz,Acc+Gyro & known&video &near-free-living  \\ 
					\midrule
					CAD\cite{10.1145/3407623}& 349$^a$ & 4,680& -& 1,107& dominant (L/R known)& Shimmers3, 15Hz, Acc+Gyro & unknown & button&  free-living  \\ 
					\midrule	
					Capture-24\cite{chan2024capture}&  151(78)$^b$ & 2,562& -& 199 & dominant (L/R unknown) & AX3,100Hz, Acc & unknown &image& free-living\\ 
					\midrule				
					Physio-ED\cite{10.3389/fnins.2019.00606}&  57 & 456 & -& 171 & two-hand & Chillband, 32Hz, Acc & unknown &schedule& in the hospital\\ 
					\hline
					\bottomrule
				\end{tabular}
				\begin{tablenotes}
					\footnotesize          
					\item[a] In \cite{10.1145/3407623}, the reported number of participants is 351, whereas the shared dataset contains 349 participants.        %
					\item[b] The dataset includes 151 participants, of whom 78 have eating-specific labels, enabling meaningful quantitative analysis.        %
				\end{tablenotes} 
			\end{threeparttable} 
		}
	\end{center}
\end{table*}

\subsection{In-meal datasets for eating gesture detection }
The in-meal datasets contain fine-grained eating gesture-level annotation, together with the FD-I dataset, they were used to train the eating gesture detection module.
\subsubsection{Meal-Only (MO) dataset}
The MO dataset \cite{10584254} consists of 46 meal sessions collected from 46 participants, containing a total of 2,894 eating gestures. Note that there is no participant overlap between the MO dataset and the FD-I datasets.
\subsubsection{OREBA dataset}
The OREBA (OREBA-DIS) dataset \cite{b51} includes 100 meal sessions obtained from 100 participants, comprising 4,790 intake gestures. The dataset was collected in Australia and involves several common eating utensil types, including forks and knives, spoons, and hand-based eating. Motion data were recorded from both wrists using two IMU wristbands.

\subsection{Annotation information}
All full-day datasets mentioned above were collected in (near-)free-living environments. However, they differ in data collection setup, sensor frequency, orientation, and annotation granularity. Among them, FD-I and CAD datasets provide time information for each eating episode. For the Capture-24 dataset, eating-specific annotations for 78 participants were provided. Eating episodes were annotated in 30 s time windows. In this study, consecutive eating windows were merged into a single episode; episodes separated by less than 30 min were further combined; and episodes shorter than 3 min were excluded. For the Physio-ED dataset, no labels were provided. However, because the data were collected in a clinical hospital, all participants adhered to standard meal and snack times: lunch at 12:15 and snacks at 10:30 and 15:00. All participants were observed during mealtime and no abnormalities were reported. We therefore assigned coarse annotations based on these time periods. An overview of the datasets used in this study is summarized in Table \ref{dataset_info}.
\subsection{Preprocessing}
The datasets varied in sampling frequency, ranging from 15 Hz to 100 Hz. To standardize the data processing pipeline, all datasets were resampled to 15 Hz using SciPy's polyphase FIR-filter-based resampling, which was considered sufficient to achieve comparable model performance based on our previous experience \cite{s5}. Additionally, the units of the accelerometer and gyroscope measurements from all datasets were converted to match the Shimmer sensors, i.e., $m/s^2$ for acceleration and $rad/s$ for angular velocity.

\section{Framework}
\subsection{Hand and orientation inference principle}
\subsubsection{General intuition}
When the wrist-worn IMU dataset lacks explicit information about sensor orientation or the hand on which the device was worn, it becomes challenging to directly apply a pretrained model. This difficulty arises from data heterogeneity: the sensor axes in the new dataset may not align with those used during model training.

\begin{figure}[t]
	\centering
	\includegraphics[scale=0.25]{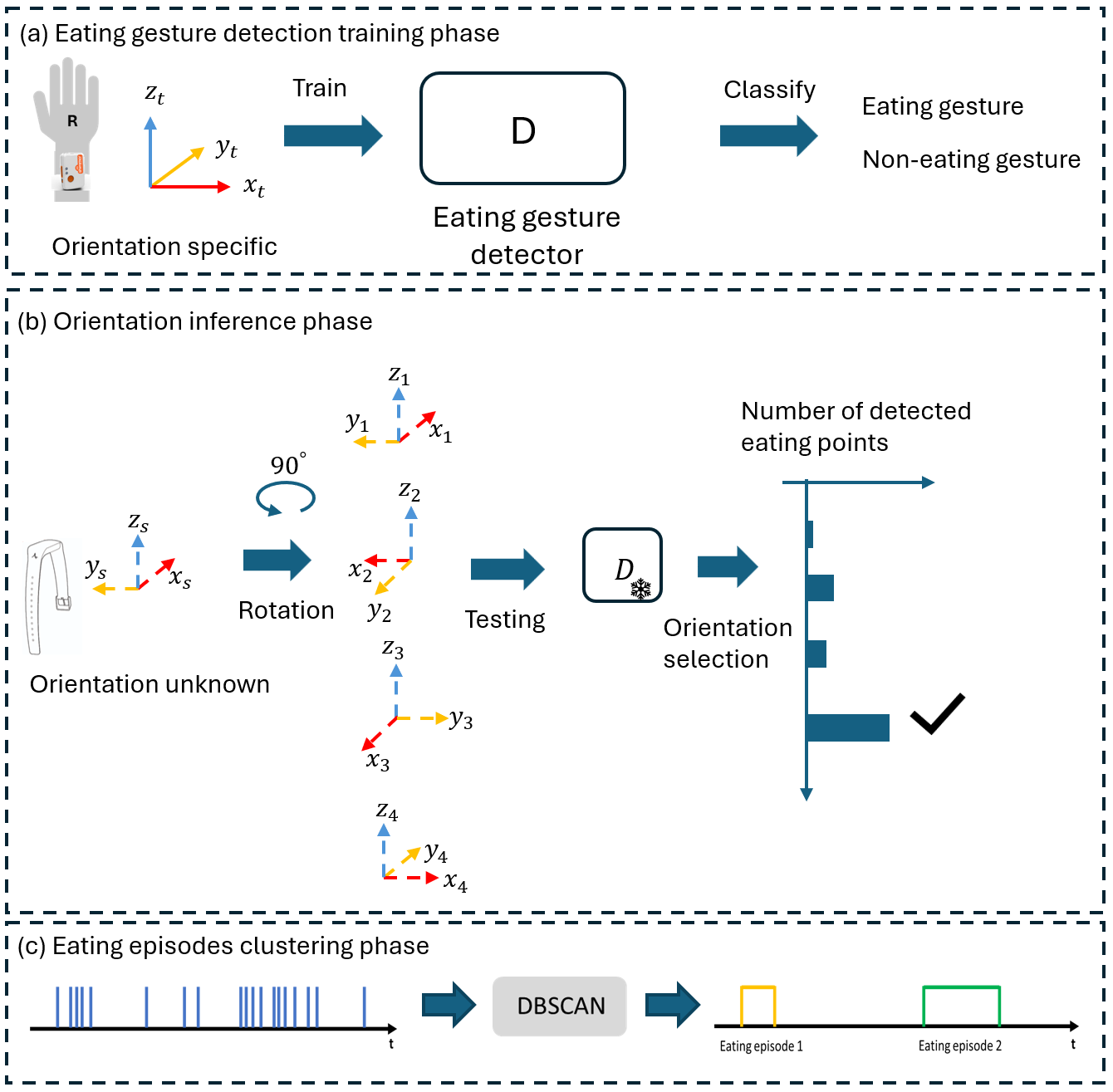}
	\vspace{-0.3cm}
	\caption{Overview of the proposed framework. (a) The eating gesture detection model is first trained using orientation-specific, fine-grained datasets. (b) The trained model is then applied to external datasets with unknown sensor placement to infer sensor orientation based on the distribution of detected eating gestures. (c) Once the orientation is determined, detected eating gestures are temporally clustered to form eating episodes.} 
	\label{framework_generic}
	\vspace{-0.3cm}
\end{figure}

We hypothesize that if the sensor orientation of the validation dataset is aligned with that of the pretrained dataset, the model will yield a higher number of predicted eating gestures. Conversely, if the orientations differ, the prediction performance will degrade, as shown in Fig. \ref{framework_generic}

To identify the correct orientation, we adopt an inverse reasoning approach. Specifically, we systematically rotate the IMU data by $90^\circ$ increments by inverting the signs of the x- and y-axes, generating four possible orientation configurations in total. Each configuration is then fed into the pretrained model, which outputs sample-wise predictions across three classes. The configuration producing the highest number of samples predicted as eating gestures is selected as the correct sensor orientation. This approach assumes that the x- and y-axes are always parallel to the plane of the sensor, while the z-axis is always perpendicular to the sensor plane. 

\subsubsection{Orientation candidates} \label{sec_oc}
To account for potential orientation mismatches between datasets, we define four candidate configurations by inverting the signs of the $x$- and $y$-axes of the IMU signals. Let the tri-axial vector be
\begin{equation}
	\mathbf{v}_s = [v_x, v_y, v_z]^T, \quad v \in \{a, g\},
\end{equation}
where $a$ represents acceleration and $g$ represents angular velocity, both defined in the sensor coordinate system.

To handle uncertainty in horizontal sensor orientation, we rotate the vector around the sensor $z$-axis by $0^\circ$, $90^\circ$, $180^\circ$, and $270^\circ$, generating four rotation candidates:
\begin{equation}
	\mathbf{v}'_i = \mathbf{R}_z(\theta_i) \, \mathbf{v}_s, \quad
	\theta_i \in \{0, \pi/2, \pi, 3\pi/2\},
\end{equation}
where the rotation matrix is
\begin{equation}
	\mathbf{R}_z(\theta_i) =
	\begin{bmatrix}
		\cos\theta_i & -\sin\theta_i & 0 \\
		\sin\theta_i &  \cos\theta_i & 0 \\
		0 & 0 & 1
	\end{bmatrix}.
\end{equation}

\subsubsection{Hand inference} \label{hand_info}
For datasets where the wearing hand is unknown, one extra step is needed to infer the hand. If the sensor is worn on the left hand, the vector is mirrored to the right-hand coordinate system using a hand-specific mirroring matrix $\mathbf{M}_v$:
\begin{equation}
	\mathbf{v}''_i = \mathbf{M}_v \, \mathbf{v}'_i,
\end{equation}
with
\begin{equation}
	\mathbf{M}_v = 
	\begin{cases}
		\mathrm{diag}([-1, 1, 1]), & v = a \text{ (acceleration)}\\[2mm]
		\mathrm{diag}([1, -1, -1]), & v = g \text{ (angular velocity)}
	\end{cases}.
\end{equation}

Assuming the wearing hand is unknown, the final candidate set for each vector is
\begin{equation}
	\mathcal{V} = \{ \mathbf{v}'_i \text{ (R-hand)},\; \mathbf{v}''_i \text{ (L-hand mirrored)} \}, \quad i = 1,2,3,4.
\end{equation}
\noindent
Thus, there are 4 rotation candidates for the right-hand assumption, 4 rotation plus mirroring candidates for the left-hand assumption, totaling 8 candidate orientations per IMU sample. This unified framework allows consistent processing of both acceleration and angular velocity across datasets with unknown sensor orientation and wearing hand.

For datasets in which the wearing hand is known (CAD and Physio-ED), the hand-mirroring operation can be directly applied to project left-hand data into a unified coordinate system prior to the orientation inference stage.

\begin{figure*}[t]
	\centering
	\includegraphics[scale=0.32]{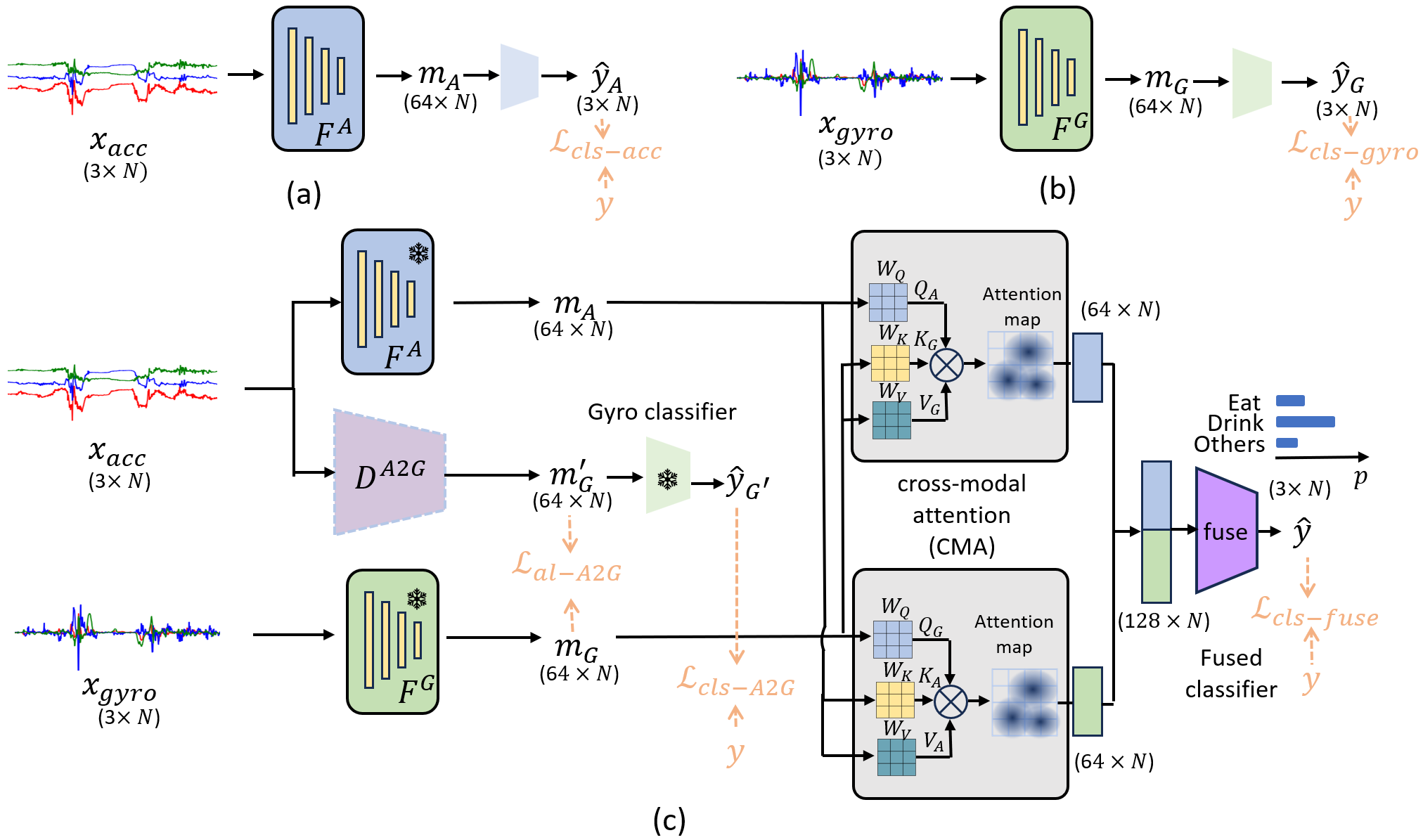}
	\vspace{-0.3cm}
	\caption{Training pipeline overview of the proposed framework: (a) training of the acc-only encoder ($\mathbb{F}^{A}$) and its classifier; (b) training of the gyro-only encoder ($\mathbb{F}^{G}$) and its classifier; (c) training of the full multi-branch architecture including acc2gyro module ($\mathbb{D}^{A2G}$), cross-modal attention (CMA), and multimodal classifier. The snowflake symbol indicates modules whose weights are frozen and inherited from (a) and (b).} 
	\label{model_train}
	\vspace{-0.3cm}
\end{figure*}

\subsection{Eating gesture detection model}
The eating gesture detection model serves as the core component for orientation inference and also forms the foundation for subsequent eating episode localization. In this study, we aim to design a flexible architecture capable of handling different input configurations, specifically both acc-only and acc+gyro data. To achieve this, acc and gyro signals are treated as two distinct branches. Accordingly, we propose a multi-branch temporal cross-attention network with virtual sensor generation (MB-TCAN-VSG), which consists of branch-specific feature encoders for independent feature extraction from acc and gyro signals, a virtual sensor generation module (acc2gyro) for reconstructing gyro features from acc data, a cross-modal attention-based fusion module for integrating complementary information from both modalities, and a final $1\times1$ Conv1d classification head.

\subsubsection{Branch-specific feature encoder}
Previous studies commonly concatenate acc and gyro signals into a single six-channel input representation and process them using a unified network (i.e., raw-data fusion). In contrast, the proposed architecture adopts separate feature encoders for each modality to enable modality-specific feature learning. For the acc branch, the encoder is defined as $\mathbb{F}^{A}: x_{acc} \rightarrow m_A$, where $x_{acc}\in\mathbb{R}^{3\times N}$ denotes the input acc signal and $m_A\in\mathbb{R}^{64\times N}$ represents the extracted intermediate feature representation. Similarly, the gyro branch employs the same encoder structure due to the identical dimensionality, producing intermediate gyro features denoted as $m_G$.

\subsubsection{Virtual sensing generation: acc2gyro}
To address scenarios where gyro measurements are unavailable, a virtual sensor generation module, referred to as acc2gyro, is introduced to reconstruct gyro representations from acc information. The mapping is formulated as $\mathbb{D}^{A2G}: x_{acc} \rightarrow m'_G$, where $m'_G\in\mathbb{R}^{64\times N}$ denotes the generated intermediate gyro features. 

\subsubsection{Cross-modal attention-based fusion}
To integrate the intermediate features extracted from the acc and gyro branches, a cross-modal attention (CMA) mechanism is employed to capture complementary information between modalities. Derived from the self-attention mechanism \cite{vaswani2017attention}, CMA differs in that the queries (Q), keys (K), and values (V) are obtained from two modalities. Specifically, queries are generated from one modality, while keys and values are generated from the other. The cross-attention from acc features to gyro features is formulated as:

\begin{equation}
	\text{crs-att}_{A \rightarrow G} = \text{softmax}\left(\frac{Q_AK_G^T}{\sqrt{d_k}}\right)V_G
\end{equation}
where $d_k$ denotes the dimension of the attention module. 

To fully leverage information from both modalities, a symmetric CMA mechanism is employed, where features are fused bidirectionally:

\begin{equation}
	m_{\text{fused}} = \text{concat}(\text{crs-att}_{A \rightarrow G}, \text{crs-att}_{G \rightarrow A})\label{eq4}
\end{equation}

\subsection{Loss function} 
\subsubsection{Sample-wise classification loss function}
\label{sec:clas}
The sample-wise classification objective consists of a classification loss and a smoothing loss. First, the cross-entropy (CE) loss is employed to optimize classification performance. Second, to reduce over-segmentation, a truncated mean squared error (T-MSE) loss computed on sample-wise log-probabilities is introduced as a smoothing regularization term. The combined loss is defined as:

\begin{equation}\mathcal{L}_{cls-fuse} =\mathcal{L}_{ce}+\lambda{\mathcal{L}_{T-MSE}}\label{eq8}\end{equation}
where $\lambda$ controls the relative contribution of the smoothing term. The $\lambda$ is set to 0.15 from \cite{b37}.

\subsubsection{A2G modality adaptation loss function}
To encourage the acc2gyro module $\mathbb{D}^{A2G}$ to generate intermediate representations that are consistent with features extracted from the actual gyro branch, a feature alignment loss is introduced. Specifically, the MSE is employed to minimize the discrepancy between reconstructed and original representations:

\begin{equation}
	\mathcal{L}_{al-A2G} = \frac{1}{n} \|  m'_G - m_G \|_2^2
	\label{eq9}\end{equation}

where $\mathcal{L}_{al-A2G}$ denotes the feature alignment loss between the original gyro features $m_G$ and the reconstructed features $m'_G$ generated by $\mathbb{D}^{A2G}$, and $n$ represents the total number of feature elements ($64\times N$). 

To further improve the discriminative capability of the reconstructed representations, an auxiliary classification loss, denoted as $\mathcal{L}_{cls-A2G}$, is introduced. Specifically, the generated features $m'_G$ are fed into a frozen gyro prediction layer, and the resulting classification loss is jointly optimized. 

The overall training objective combines the primary classification loss and the auxiliary modality adaptation loss, and is defined as:

\begin{equation}\mathcal{L}_{\text{total}} =  \alpha\mathcal{L}_{al-A2G}+\mathcal{L}_{cls-A2G}+\mathcal{L}_{\text{cls-fuse}}\label{eq10}\end{equation}

where $\alpha$ controls the relative contribution of the alignment loss and is set to 0.1, based on empirical evaluation.

\subsection{Training and testing}  
\label{sec:tran}
A two-step training strategy is adopted. In the first stage, separate uni-branch models are independently trained for acc and gyro data, as illustrated in Fig. \ref{model_train}(a) and (b). Each uni-branch model comprises two components: a feature encoder that extracts intermediate representations and a classifier that generates class probabilities from these representations. In the second stage, the pre-trained acc and gyro encoders, together with the gyro classifier, are incorporated into the multi-branch learning framework with their parameters frozen. The remaining components, including the acc2gyro module, CMA, and multimodal classifier, are then optimized within the integrated framework, as shown in Fig. \ref{model_train}(c).

\begin{figure}[t]
	\centering
	\includegraphics[scale=0.22]{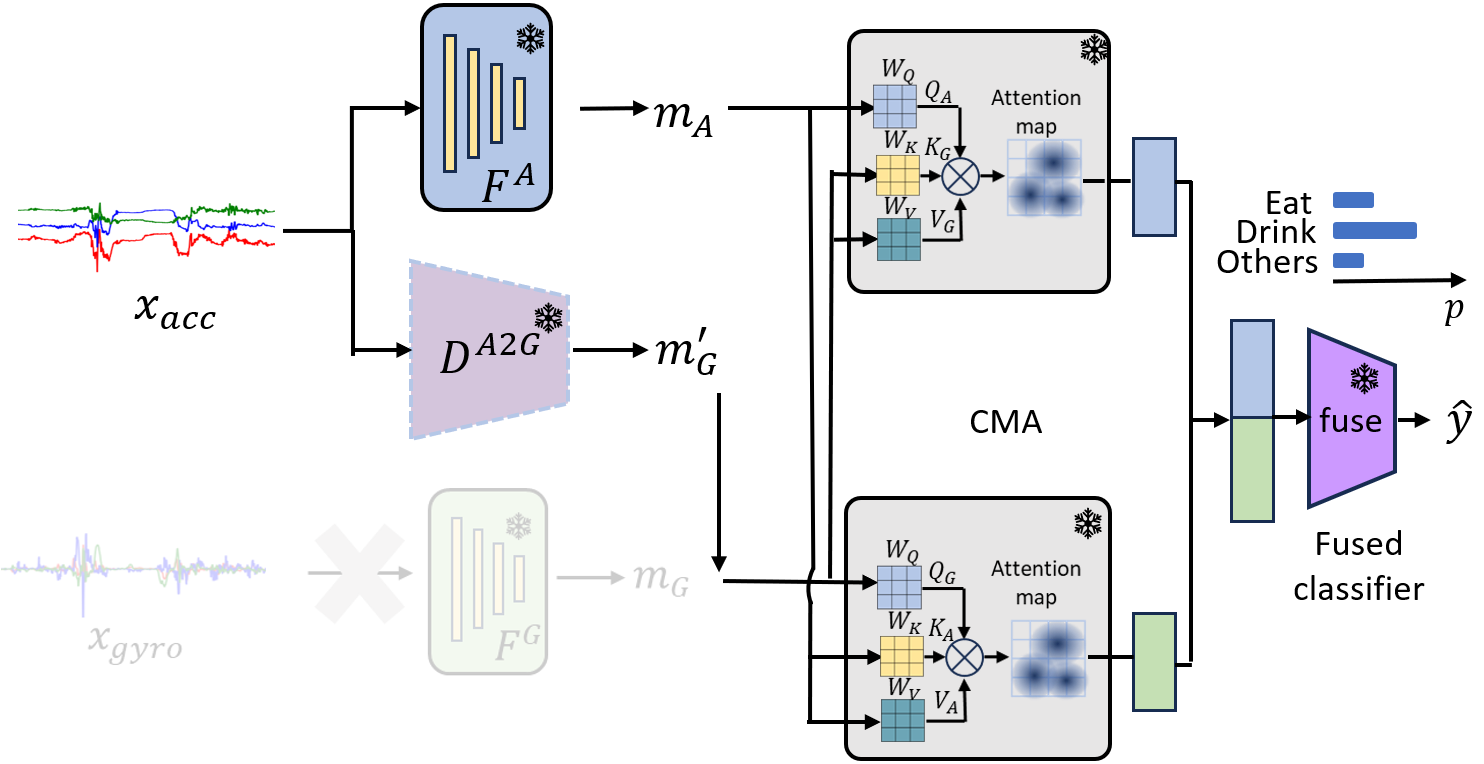}
	\vspace{-0.3cm}
	\caption{The inference pipeline for acc-only scenario. When gyro data is not available, the acc2gyro module ($\mathbb{D}^{A2G}$) generates surrogate gyroscope representations, which are then used in place of the missing gyro signals for downstream inference.} 
	\label{model_inference_acc_only}
	\vspace{-0.3cm}
\end{figure}

During inference, under the acc+gyro setting, the extracted features $m_A$ and $m_G$ are fed into the CMA module for feature fusion. Under the acc-only setting, as illustrated in Fig. \ref{model_inference_acc_only}, the outputs of $\mathbb{F}^A$ and $\mathbb{D}^{A2G}$, namely $m_A$ and the generated feature representation $m_G'$, respectively, are used as inputs to the CMA module for feature fusion.

\subsection{Model implementation}
Building upon our previous work \cite{10584254}, the one-dimensional temporal convolutional neural network (1D-TCN) is adopted as the feature encoder for both the acc and gyro branches. The TCN is selected for its ability to efficiently capture long-range temporal dependencies through stacked dilated convolutional layers, while also enabling multi-scale temporal modeling, as demonstrated in \cite{8099596}. To reduce the overall complexity of the framework, the same 1D-TCN architecture is additionally employed as the acc2gyro module. Within the CMA module, eight attention heads are used, each with a dimensionality of 8. The model is trained using the Adam optimizer with a learning rate of 0.0005 for 100 epochs. The input window length is set to 60 seconds, and the batch size is set to 4. All training, validation, and testing experiments were conducted on an NVIDIA V100-SXM2-32GB GPU provided by the Vlaams Supercomputer Centrum (VSC)\footnote{See https://www.vscentrum.be/}.

\subsection{Orientation selection}
Once the model is trained, external datasets with unknown orientation are fed into the trained model with different orientation candidates. Each rotated signal $\mathbf{v}'_i$ is then fed into the pretrained
three-class classifier $f(\cdot)$ to produce sample-wise predictions:
\begin{equation}
	\hat{y}_{i,t} = f(\mathbf{v}'_{i,t}),
	\quad \hat{y}_{i,t} \in \{\text{eating}, \text{drinking}, \text{other}\},
\end{equation}
where $t$ indexes the time samples.

For each orientation $i$, we compute the total number of samples
predicted as the \emph{eating} class:
\begin{equation}
	N_{i}^{(\text{eat})} =
	\sum_{t=1}^{T} \mathbb{I}\big(\hat{y}_{i,t} = \text{eating}\big),
\end{equation}
where $\mathbb{I}(\cdot)$ denotes the indicator function and $T$ is the
sequence length.

The orientation producing the largest number of eating predictions is
selected as the correct configuration:
\begin{equation}
	i^{*} = \arg\max_{i} N_{i}^{(\text{eat})}.
\end{equation}
The corresponding output sequence $\hat{y}_{i^{*},t}$ is then adopted
as the final prediction for the dataset. The overall pipeline is shown in Algorithm \ref{alg:orientation_inference}.

\begin{algorithm}[t]
	\caption{Task-Driven Orientation Inference}
	\label{alg:orientation_inference}
	\begin{algorithmic}[1]
		\STATE \textbf{Input:} Raw IMU data $\mathbf{v}_s$, pretrained model $f(\cdot)$
		\STATE \textbf{Output:} Correct orientation index $i^{*}$ and predictions $\hat{y}_{i^{*},t}$
		\STATE Define $\Theta = \{0, \tfrac{\pi}{2}, \pi, \tfrac{3\pi}{2}\}$
		\FOR{each $\theta_i \in \Theta$}
		\STATE $\mathbf{v}'_{i} = \mathbf{R}_z(\theta_i)\, \mathbf{v}_s$
		\STATE $\hat{y}_{i,t} = f(\mathbf{v}'_{i,t})$
		\STATE $N_{i}^{(\text{eat})} = \sum_{t=1}^{T} \mathbb{I}(\hat{y}_{i,t} = \text{eating})$
		\ENDFOR
		\STATE $i^{*} = \arg\max_{i} N_{i}^{(\text{eat})}$
		\STATE \textbf{return} $i^{*}$ and $\hat{y}_{i^{*},t}$
	\end{algorithmic}
\end{algorithm}

\begin{figure}[b]
	\centering
	\includegraphics[scale=0.27]{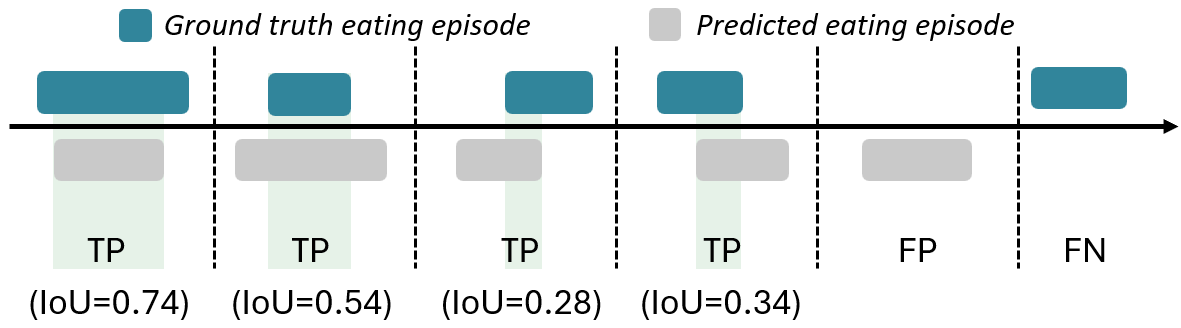}
	\vspace{-0.3cm}
	\caption{Examples illustrating the evaluation of eating episode detection from \cite{sharma2022top, tang2023detecting}. A detected eating episode is counted as TP if it overlaps with a ground truth meal; otherwise, it is classified as FP. Ground truth meals without overlapping detections are counted as FN.} 
	\label{seg_eva}
	\vspace{-0.3cm}
\end{figure}

\subsection{Eating episode recognition}
After determining the orientation, the predicted sequence of eating gestures is further utilized to identify eating episodes. Since eating episodes are primarily characterized by eating gestures, any detected drinking gestures are excluded at this stage. The remaining predicted gestures are then grouped into eating episodes using the 1D-DBSCAN clustering algorithm \cite{s42}. DBSCAN forms clusters according to the density of data points. In our implementation, the \emph{sklearn.cluster.DBSCAN} function is adopted with the epsilon parameter set to 3 minutes and the minimum number of samples set to 5 based on previous study \cite{10584254}. Through this clustering process, isolated bite events with low density are treated as noise, while temporally dense bite sequences are grouped into clusters representing eating episodes, as illustrated in Fig. \ref{framework_generic}(c). After clustering, additional post-processing steps similar to those described in \cite{m1, s35} are applied to refine the detected episodes. Specifically, two adjacent episodes are merged if the time interval between them is shorter than 30 minutes. Moreover, if the duration of a resulting episode is less than 3 minutes after merging, that episode is discarded.

\section{Validation and Evaluation Scheme}
\subsection{Validation strategy}
\subsubsection{Internal validation}
To assess the effectiveness of the proposed multi-branch model for sample-wise classification, as well as the resulting gesture-wise and episode-wise detection performance, a seven-fold cross-validation strategy was applied to the FD-I dataset. Specifically, in each fold, all corresponding IMU data from participants assigned to the test partition were exclusively used for testing to prevent information leakage. In addition, the MO and OREBA datasets were included in the training set for every fold. 
\subsubsection{External validation}
According to the main goal of the study, we evaluate the model's ability to generalize across heterogeneous data sources. The model is initially trained on a composite dataset comprising FD-I, OREBA, and MO. The model is subsequently evaluated on three independent, hold-out datasets: CAD, Capture-24, and Physio-ED. This ensures that the performance metrics reflect the model's robustness to unseen sensor characteristics and demographic variations.

\subsection{Evaluation}
\subsubsection{Gesture evaluation}
For internal validation, gesture-level performance can be evaluated using the available eating gesture annotations. Since the model outputs sample-wise multi-class predictions and the dataset is inherently imbalanced, Cohen's kappa is employed to assess sample-wise classification performance. However, the ultimate goal of bite detection is accurate bite counting rather than sample-level classification. Therefore, we additionally evaluate detection performance using the segment-wise F1-score, as adopted in prior work \cite{wang2024evaluation}. The intersection over union (IoU) is first calculated between each predicted eating gesture segment and the corresponding ground-truth (GT) segment. Predicted segments with an IoU greater than 0.5 are considered true positives (TP), while unmatched GT and predicted segments are counted as false negatives (FN) and false positives (FP), respectively. The segment-wise F1-score is then computed from these values.

\subsubsection{Episode evaluation}
Episode-level metrics are computed by comparing the temporal overlap between the eating episodes detected by the models and the GT episodes, as illustrated in Fig. \ref{seg_eva}. A predicted episode is considered a TP if it overlaps with any GT eating episode. Conversely, an FP occurs when the model predicts an eating episode that does not correspond to any GT episode. A FN is recorded when a ground-truth eating episode is not detected by the model. Subsequently, the episode-wise F1-score is adopted as the primary evaluation metric. To ensure a fair comparison with prior studies \cite{sharma2022top, tang2023detecting}, we follow their evaluation protocol, where IoU $> 0$ is used for eating episode detection (TP definition), while our eating gesture detection typically uses IoU $> 0.5$. In addition to detection performance, the averaged IoU score on TP episodes is evaluated for datasets with segment annotations (CAD and Capture-24).

\subsection{Benchmark models}
\subsubsection{CNN-LSTM}
The CNN-LSTM comprises three 1D convolutional layers followed by a single LSTM layer. The CNN block includes three layers, each with 64 kernels of size $3$, and utilizes the ReLU activation function. The output from the CNN is flattened within each time step and subsequently fed into the LSTM layer, which contains 64 hidden units. 
\subsubsection{ResNet-LSTM}
The ResNet-LSTM architecture integrates a one-dimensional Residual Network (1D-ResNet-10) backbone with a subsequent LSTM layer, as described in \cite{m13}. ReLU is employed after each convolutional layer. The output from the ResNet backbone is flattened and subsequently input into the LSTM layer, which contains 128 hidden units. 
\subsubsection{DCNN-LSTM-MHA}
The DCNN-LSTM-MHA integrates a three-layer 1D-CNN with 32,64,64 filters, three LSTM layers with hidden sizes of 64, respectively, and two 8-head attention modules with a dimension of 64 as described in \cite{9781408}.
\subsubsection{TCCSNet}
The temporal-channel convolution with self-attention network (TCCSNet) architecture \cite{ESSA2023110867} integrates two parallel branches: a time-wise branch and a channel-wise branch. Each branch comprises a two-layer 1D-CNN with 32 and 64 filters, respectively, followed by two layers of 8-head self-attention modules with an embedding dimension of 64.
\subsubsection{TCN-MHA}
The TCN with a multi-head attention mechanism (TCN-MHA), adapted from our previous work \cite{10584254}, is included as a benchmark model. The architecture consists of a 9-layer 1D-TCN, with 64 filters in each layer for temporal feature extraction. The extracted features are subsequently processed by an MHA module comprising 8 attention heads, each with a dimensionality of 8.
\subsubsection{HARNet-10s}
Recently, self-supervised foundation models (FMs), which are pretrained on large-scale unlabeled datasets and subsequently fine-tuned on smaller labeled datasets for downstream tasks, have attracted considerable attention. In this study, we adopt the acc-based HAR foundation model proposed by Yuan et al. \cite{yuan2024self} and fine-tune it using the training data. Before fine-tuning, the datasets are preprocessed to satisfy the model's input requirements. Based on preliminary experiments, the number of fine-tuning epochs is set to two. As HARNet-10s is designed exclusively for acc data, it is evaluated only in the acc-only experiments.

\begin{table*}[t]
	\caption{Internal validation on FD-I dataset}
	\vspace{-0.2cm}
	\label{results_fd1}
	\begin{center}
		\scalebox{0.8}{
			\begin{tabular}{l|l|c|c|c|c}
				\toprule
				\hline
				\rule{0pt}{3ex}Modality&Model&Kappa score &Gesture F1-score&Episode F1-score&Episode IoU \\ 
				\midrule
				\multirow{2}*{Acc+Gyro}&TCN &0.673&0.726&0.920& 0.838  \\ 
				~&MB-TCAN&0.713&0.732&0.943& 0.854 \\ 
				\midrule
				\multirow{2}*{Acc-only}&TCN &0.608&0.616&0.887& 0.783  \\ 
				~&MB-TCAN-VSG &0.670&0.665&0.887& 0.808  \\
				
				\hline
				\bottomrule
		\end{tabular}}
	\end{center}
	\vspace{-0.3cm}
\end{table*}

\begin{table*}[t]
	\caption{External validation performance for eating episode detection}
	\vspace{-0.2cm}
	\label{data_eva}
	\begin{center}
		\scalebox{0.8}{
			
			\begin{tabular}{l|l|l|l|l|l|l|l|l|l}
				\toprule
				\hline				
				& & & & & & & \multicolumn{2}{c}{F1-score}& \\
				Dataset & Model & \#TP & \#FP & \#FN & Pre & Rec & Overall &Participant&IoU\\
				&&&&&&&& (mean $\pm$ SD) & \\
				\midrule
				\multirow{6}*{CAD}&CNN-LSTM& 801&317&306& 0.716 & 0.724 & 0.720&0.698 $\pm$ 0.255 & 0.552 \\ 
				~&ResNet-LSTM&327  & \textbf{63}& 780&\textbf{0.838} & 0.295&0.437& 0.371 $\pm$ 0.336& 0.529\\
				~&DCNN-LSTM-MHA& 575 & 138& 532& 0.806& 0.519&0.632& 0.575 $\pm$ 0.331 & 0.564\\
				~&TCCSNet&699  & 199& 408&0.778 &0.631 &0.697& 0.640 $\pm$ 0.331 &0.574 \\
				(acc+gyro)&TCN-MHA&  \textbf{894}&418&\textbf{213}& 0.681 & \textbf{0.808} & 0.739& 0.703 $\pm$ 0.280 & 0.548\\
				~&MB-TCAN&  801&225&306& 0.781  & 0.724 & \textbf{0.751}& \textbf{0.727 $\pm$ 0.266} & \textbf{0.591}\\ 
				\midrule
				\multirow{7}*{CAD}&CNN-LSTM& 699&191&408& 0.785 & 0.631 & 0.700& 0.667 $\pm$ 0.291& 0.543 \\ 
				~&ResNet-LSTM& 786 &394 &321 &0.667 &0.710 &0.687& 0.671 $\pm$ 0.263 &0.501 \\
				~&DCNN-LSTM-MHA&725  & 267&382 &0.731 &0.655 &0.691& 0.662 $\pm$ 0.278 &0.563 \\
				~&TCCSNet&491  &\textbf{102} &616 &\textbf{0.828} &0.444 &0.578& 0.523 $\pm$ 0.332 &\textbf{0.574} \\
				(acc-only)&HARNet-10s& 578&158 & 529 &0.785  & 0.522 & 0.627 & 0.585 $\pm$ 0.307 & 0.515  \\ 
				~&TCN-MHA&  \textbf{833}&447&\textbf{274}& 0.651 & \textbf{0.752} & 0.700& 0.692 $\pm$ 0.245 &  0.529 \\ 
				~&MB-TCAN-VSG &  798&303&309& 0.725  & 0.721 & \textbf{0.723}& \textbf{0.706 $\pm$ 0.262} &  0.569\\ 
				\midrule				
				\multirow{7}*{Capture-24}&CNN-LSTM& 129&134&70& 0.490 & 0.648 & 0.558& 0.522 $\pm$ 0.291 & 0.524 \\ 
				~&ResNet-LSTM& 137 & 205 & 62 & 0.401 & 0.688 & 0.506& 0.481 $\pm$ 0.252 & 0.459 \\
				~&DCNN-LSTM-MHA& 126 &150 &73 &0.457 &0.633 &0.531& 0.509 $\pm$ 0.270 & 0.521\\
				~&TCCSNet&88  & 78& 111&0.530 &0.442 &0.482& 0.433 $\pm$ 0.302 &\textbf{0.622} \\
				(acc-only)&HARNet-10s& 89 &\textbf{56} & 110 & \textbf{0.614} & 0.447 & 0.517 & 0.440 $\pm$ 0.343 & 0.559  \\ 
				~&TCN-MHA&  \textbf{145}&186&\textbf{54}& 0.438 & \textbf{0.729} & 0.547& 0.534 $\pm$ 0.259  &  0.563 \\ 
				~&MB-TCAN-VSG &  142&139&57& 0.505  & 0.714 & \textbf{0.592}& \textbf{0.568 $\pm$ 0.260}  &  0.551\\ 
				\midrule
				\multirow{7}*{Physio-ED}&CNN-LSTM& 133&46&38& 0.743 & 0.778& 0.760 & 0.734 $\pm$ 0.266 &- \\ 
				~&ResNet-LSTM&146  &60 & 25&0.709 &0.854 & 0.775& 0.763 $\pm$ 0.206 & - \\
				~&DCNN-LSTM-MHA& 133 &57 &38 &0.700 &0.778 &0.737& 0.717 $\pm$ 0.243 & -\\
				~&TCCSNet&118  & 27& 53& 0.814&0.690 & 0.747& 0.710 $\pm$ 0.275& - \\				
				(acc-only)&HARNet-10s& 100 & \textbf{18} & 71 &\textbf{0.847}  & 0.585 & 0.692 & 0.625 $\pm$ 0.319 &  - \\ 
				~&TCN-MHA&  135&45&36& 0.750  & 0.789 & 0.769 & 0.755 $\pm$ 0.251&  -\\ 
				~&MB-TCAN-VSG &  \textbf{146}&51&\textbf{25}& 0.741 & \textbf{0.854} & \textbf{0.793}& \textbf{0.779 $\pm$ 0.214} & -\\  
				\hline
				\bottomrule
		\end{tabular}}
	\end{center}
	\vspace{-0.3cm}
\end{table*}

\begin{figure*}[t]
	\centering
	\includegraphics[scale=0.35]{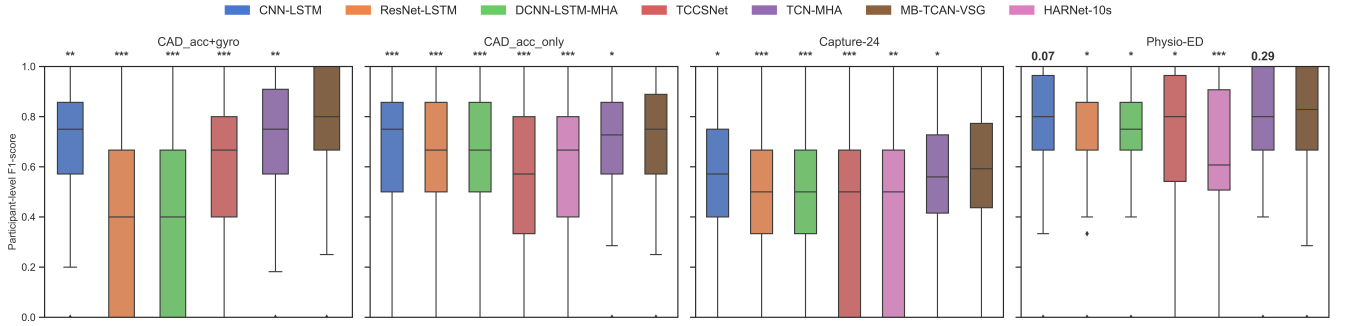}
	\vspace{-0.3cm}
	\caption{Boxplots of individual eating episode detection F1-score distribution using different models on different datasets. The $p$ values  after Benjamini–Hochberg (BH) multiple comparison correction are shown on top of the box plots. $p<0.05$: *,$p<0.01$: **,$p<0.001$: ***. Note that the relatively wide dispersion in participant-level F1-scores for certain datasets stems from the inherent sparsity of eating episodes per individual, where a single misclassification can disproportionately affect individual metrics.} 
	\label{indv_boxplot}
	\vspace{-0.3cm}
\end{figure*}

\section{Results}
\subsection{Internal validation performance}
The internal validation experiment was conducted to evaluate the effectiveness of the proposed multi-branch architecture and the acc2gyro module at both the sample-wise and gesture-wise levels. The results on FD-I are presented in Table \ref{results_fd1}. In the acc+gyro setting, the proposed MB-TCAN (without acc2gyro) consistently outperformed vanilla TCN (the backbone of MB-TCAN), demonstrating the advantage of modality-specific feature extraction over directly concatenating acc and gyro signals as a single input modality. Specifically, MB-TCAN achieved a higher sample-wise Cohen's kappa score (0.713 vs. 0.673) and gesture-wise F1-score (0.732 vs. 0.726). These improvements further translated into enhanced episode detection performance, yielding a higher episode-wise F1-score (0.943 vs. 0.920) and IoU score (0.854 vs. 0.838).

When gyro data were unavailable, incorporating the proposed acc2gyro module substantially improved performance compared with the acc-only baseline. The sample-wise kappa score increased from 0.608 to 0.670, while the gesture-wise F1-score improved from 0.616 to 0.665. Although both approaches achieved the same episode-wise F1-score, the acc2gyro-enhanced model attained a higher IoU score (0.808 vs. 0.783), indicating more accurate temporal localization of eating episodes.

\subsection{External validation performance}
The external validation results on the CAD, Capture-24, and Physio-ED datasets are summarized in Table \ref{data_eva}. For the CAD dataset, both acc+gyro and acc-only scenarios were evaluated. In the acc+gyro setting, the proposed MB-TCAN-VSG model achieved the highest episode-wise overall F1-score (0.751) and IoU score (0.591), outperforming TCN-MHA, which obtained an F1-score of 0.739 and an IoU of 0.548. Although MB-TCAN-VSG detected fewer TP episodes than TCN-MHA (\#TP: 801 vs. 894), it substantially reduced the number of FPs (225 vs. 418), resulting in a considerably lower FP/TP ratio (0.281 vs. 0.468). This suggests that the proposed model provides more reliable episode detection by reducing false alarms while maintaining competitive sensitivity. In the acc-only setting, MB-TCAN-VSG continued to outperform the TCN-based baseline, achieving an F1-score of 0.723 and an IoU of 0.569, compared with 0.700 and 0.519, respectively. The performance of ResNet-LSTM in the acc+gyro setting is significantly lower than that in the acc-only setting. This degradation is not observed in CNN-LSTM or TCN-based models, suggesting that the robustness is not determined by the presence of recurrent modeling, but rather by the extent to which different architectures suppress gyroscope-induced domain-specific variations during feature extraction.

The Capture-24 dataset proved to be the most challenging benchmark. The proposed MB-TCAN-VSG model achieved the highest episode-wise F1-score of 0.592, with an IoU of 0.551. The relatively modest performance was primarily associated with lower precision (0.505), indicating a larger number of FP detections. CNN-LSTM achieved the second-highest F1-score (0.558). While TCCSNet attained the highest IoU score (0.622), its F1-score is substantially lower(0.482). For Physio-ED dataset, the MB-TCAN-VSG model again achieved the best overall performance, obtaining the highest episode-wise F1-score of 0.793. The second-best result was achieved by ResNet-LSTM, with an F1-score of 0.775. 

The participant-level F1-score distributions exhibit a similar trend to the overall F1-scores reported in Table \ref{data_eva}. The detailed boxplots of participant-level distributions, together with the statistical significance analysis, are depicted in Fig. \ref{indv_boxplot}. The proposed model significantly outperforms all alternative models on CAD (acc+gyro, acc-only), and Capture-24. However, on Physio-ED, its performance does not significantly differ from CNN-LSTM ($p=0.07$) and TCN-MHA ($p=0.29$).

\subsection{Task-driven orientation selection analysis}
To evaluate the sensitivity of eating episode detection to sensor placement, we conducted a comparative analysis of four canonical axis-aligned orientations against our proposed per-participant orientation inference method. As summarized in Table \ref{compare_orient}, individual orientations exhibited extreme performance variability, with Orientations 1, 3, and 4 failing to exceed an F1-score of 0.1 and suffering from high FN ($>$1000). While Orientation 2 emerged as the most viable placement with an F1-score of 0.698 and 702 TP, which implies that the majority of participants in the CAD dataset wear the sensor in Orientation 2. By enabling participant-wise orientation inference through the proposed approach, the method achieves a higher F1-score of 0.751 and reduces the number of missed events to 306. These results underscore the strong orientation sensitivity of eating gesture detection and demonstrate the feasibility and effectiveness of the proposed approach for sensor orientation inference.

\begin{table}[t]
	\caption{External validation performance on CAD dataset with different orientations}
	\vspace{-0.2cm}
	\label{compare_orient}
	\begin{center}
		\scalebox{0.9}{
			\begin{threeparttable}
				\begin{tabular}{l|l|l|l|l|l|l}
					\toprule
					\hline
					\rule{0pt}{3ex}Orientations&\#TP &\#FP&\#FN &Pre & Rec & F1-score \\ 
					\midrule
					1 &30&7&1077& 0.811 & 0.027 & 0.052  \\ 
					2 &702&203&405&0.776 & 0.634 & 0.698 \\
					3 &81&28&1026& 0.743 & 0.073 & 0.013 \\
					4 &10&5&1097& 0.667 & 0.009 & 0.018 \\
					SELC$^a$ &801&225&306& 0.781 & 0.724 & 0.751 \\
					\bottomrule
				\end{tabular}
				\begin{tablenotes}    
					\footnotesize              
					\item[a] SELC represents the proposed per-participant orientation inference method.   %
				\end{tablenotes}
			\end{threeparttable}
		}
	\end{center}
	\vspace{-0.3cm}
\end{table}

\begin{table}[t]
	\caption{Performance comparison between orientation-inference and data augmentation paradigm on acc-only scenario}
	\vspace{-0.2cm}
	\label{compare_paratigm}
	\begin{center}
		\scalebox{0.87}{
			
			\begin{tabular}{l|l|l|l|l|l|l|l}
				\toprule
				\hline
				\rule{0pt}{3ex}Paradigm&Dataset&\#TP &\#FP&\#FN &Pre & Rec & F1-score \\ 
				\midrule
				\multirow{3}*{Orientation-}&CAD& 833&447&274& 0.651 & 0.752 & 0.700 \\ 
				~&Capture-24& 145&186&54& 0.438 & 0.729 & 0.547 \\ 
				Inference&Physio-ED& 135&45&36& 0.750 & 0.789 & 0.769 \\
				\midrule
				\multirow{3}*{Data}&CAD& 797&372&310& 0.682 & 0.720 & 0.700 \\ 
				~&Capture-24& 95&83&104& 0.534 & 0.477 & 0.504 \\ 
				Augmentation&Physio-ED& 123&46&48& 0.728& 0.719& 0.724 \\ 
				\hline
				\bottomrule
		\end{tabular}}
	\end{center}
	\vspace{-0.3cm}
\end{table}

\subsection{Data augmentation-based paradigm}
Apart from the proposed orientation inference step, an alternative strategy is to augment the training data with rotated versions of the sensor signals and train the model directly on the augmented dataset. Under this paradigm, the model is expected to be more robust to orientation variations, allowing its predictions to be used directly during external validation without orientation inference. The performance of the two paradigms is compared in Table~\ref{compare_paratigm}. To isolate the effect of the orientation handling strategy and avoid potential confounding effects introduced by the acc2gyro module, the acc-only TCN-MHA model was used for this comparison. For the CAD dataset, the data augmentation paradigm achieved the same F1-score as the orientation inference approach (0.700). However, on the Capture-24 and Physio-ED datasets, the orientation inference paradigm consistently outperformed the data augmentation approach, yielding F1-scores of 0.547 vs. 0.504 and 0.769 vs. 0.724, respectively. These results suggest that inferring sensor orientation during external inference is more effective than relying solely on rotation-based data augmentation to improve robustness against orientation variability.

\begin{table}[t]
	\caption{Computational complexity analysis of models}
	\label{comtutation_analysis}
	\begin{center}
		\scalebox{0.85}{
			\begin{threeparttable} 
				\begin{tabular}{l|llll}
					\toprule
					\hline
					\multirow{2}*{Model}&\#Params&\#FLOPs&Memory&Latency\\
					~& &(M)&(MB)&(ms)\\
					\midrule
					CNN-LSTM       & 58.8K   & 56.8   & 2.24    & 5.73   \\
					ResNet-LSTM    & 648.8K  & 620.0  & 14.91   & 15.29  \\
					DCNN-LSTM-MHA  & 75.5K   & 56.8   & 2.24    & 33.20   \\
					TCCSNet        & 46.9K   & 28.8   & 2.19    & 29.61 \\
					HARNet-10s     & 10.52M  & 288.0  & 47.37   & 10.44  \\
					TCN-MHA        & 169.9K  & 146.0  & 10.48   & 34.06  \\
					MB-TCAN-VSG    & 340.1K  & 292.0  & 20.97   & 70.99  \\ 
					\hline
					\bottomrule
				\end{tabular} 
				\begin{tablenotes}    
					\footnotesize              
					\item[a] The latency represents the inference time needed for processing 1 min acc-only data.%
				\end{tablenotes}
			\end{threeparttable} 
		}
	\end{center}
\end{table}

\begin{table}[t]
	\caption{Comparison to existing eating episode detection studies on CAD dataset}
	\vspace{-0.2cm}
	\label{compare_litera}
	\begin{center}
		\scalebox{0.8}{
			\begin{tabular}{l|l|l|l|l}
				\toprule
				\hline
				\rule{0pt}{3ex}Work & Approach  & Validation &F1-score& FP/TP \\ 
				\midrule
				
				Sharma et al (2020) \cite{10.1145/3407623}  & \multirow{4}*{Top-down} & \multirow{4}*{Internal;5-fold} & 0.272& 5.2\\ 
				Sharma et al (2022) \cite{sharma2022top}& ~ & ~ &  0.534&1.7\\ 
				Dong et al (2014) \cite{s30, tang2023detecting}&   ~ & ~ & 0.336&3.8 \\ 
				Tang et al (2024) \cite{tang2023detecting} &  ~ & ~&  0.568&1.4\\ 
				\midrule
				Proposed approach&  Bottom-up & External & 0.751&0.28\\ 
				\hline
				\bottomrule
		\end{tabular}}
	\end{center}
	\vspace{-0.3cm}
\end{table}

\section{Discussion}
\subsection{Eating episode detection framework} 
Existing eating episode detection approaches are predominantly validated using internal evaluation schemes. Such evaluation settings limit the assessment of model robustness and constrain applicability in real-world scenarios, where variations in populations, sensor configurations, and recording conditions are inevitable. We introduce a sensor orientation inference pipeline based on an eating gesture detection model trained on the MO, OREBA, and FD-I datasets. The results in Table \ref{compare_orient} demonstrate that eating gesture detection performance is highly sensitive to sensor orientation. Rather than treating this sensitivity as a limitation, the proposed pipeline explicitly exploits it to infer both wearing hand and sensor orientation prior to eating episode detection.

The fusion of acc and gyro modalities is widely adopted in automated food intake monitoring because the two modalities provide complementary information regarding linear and rotational wrist movements. Comparative results with and without gyro input on FD-I (Table \ref{results_fd1}) and CAD (Table \ref{data_eva}) highlight the importance of gyro information for this task. However, in free-living and long-term deployment scenarios, gyroscope sensing is often constrained by its relatively high power consumption. Unlike previous studies, this work proposes a multi-branch architecture that processes acc and gyro signals independently and incorporates an acc2gyro module for virtual gyro feature generation. This design provides several advantages. First, it supports both acc+gyro and acc-only scenarios, improving model flexibility and practical usability. Second, compared with conventional early-fusion approaches that directly combine raw acc and gyro signals, the multi-branch strategy achieves superior performance. One possible explanation is that independent processing allows each branch to learn modality-specific representations more effectively before feature interaction occurs. Third, the integration of the acc2gyro module further improves performance by generating virtual gyros representations from acc signals, partially compensating for missing gyro information.

The computational efficiency of the proposed models was assessed, as shown in Table \ref{comtutation_analysis}. Although the proposed method achieves competitive performance, it maintains a relatively compact model size (340.1K parameters) compared with heavier baselines such as ResNet-LSTM (648.8K) and HARNet-10s (10.52M). Similarly, its computational cost (292.0 MFLOPs) remains considerably lower than that of more resource-intensive architectures, including ResNet-BiLSTM (620.0 MFLOPs). Inference latency was measured using a 1-minute input sequence on a laptop equipped with an Intel Core i7-10750H CPU @ 2.6 GHz (6 cores, without GPU acceleration). The MB-TCAN-VSG model required 70.99 ms to generate predictions in the acc-only setting. Although this latency was higher than that of the other baselines, the difference is primarily attributable to the additional bidirectional CMA module and multi-branch feature encoders.

In addition, the use of HARNet-10s, a pretrained large foundation model with fine-tuning, a widely adopted paradigm in the era of large-scale FM, was also investigated. Although the obtained performance remains lower than that of the proposed framework, it required only two fine-tuning epochs to achieve competitive results, highlighting the potential of FM-based approaches for efficient eating behavior recognition.

\begin{figure*}[t]
	\centering
	\includegraphics[scale=0.25]{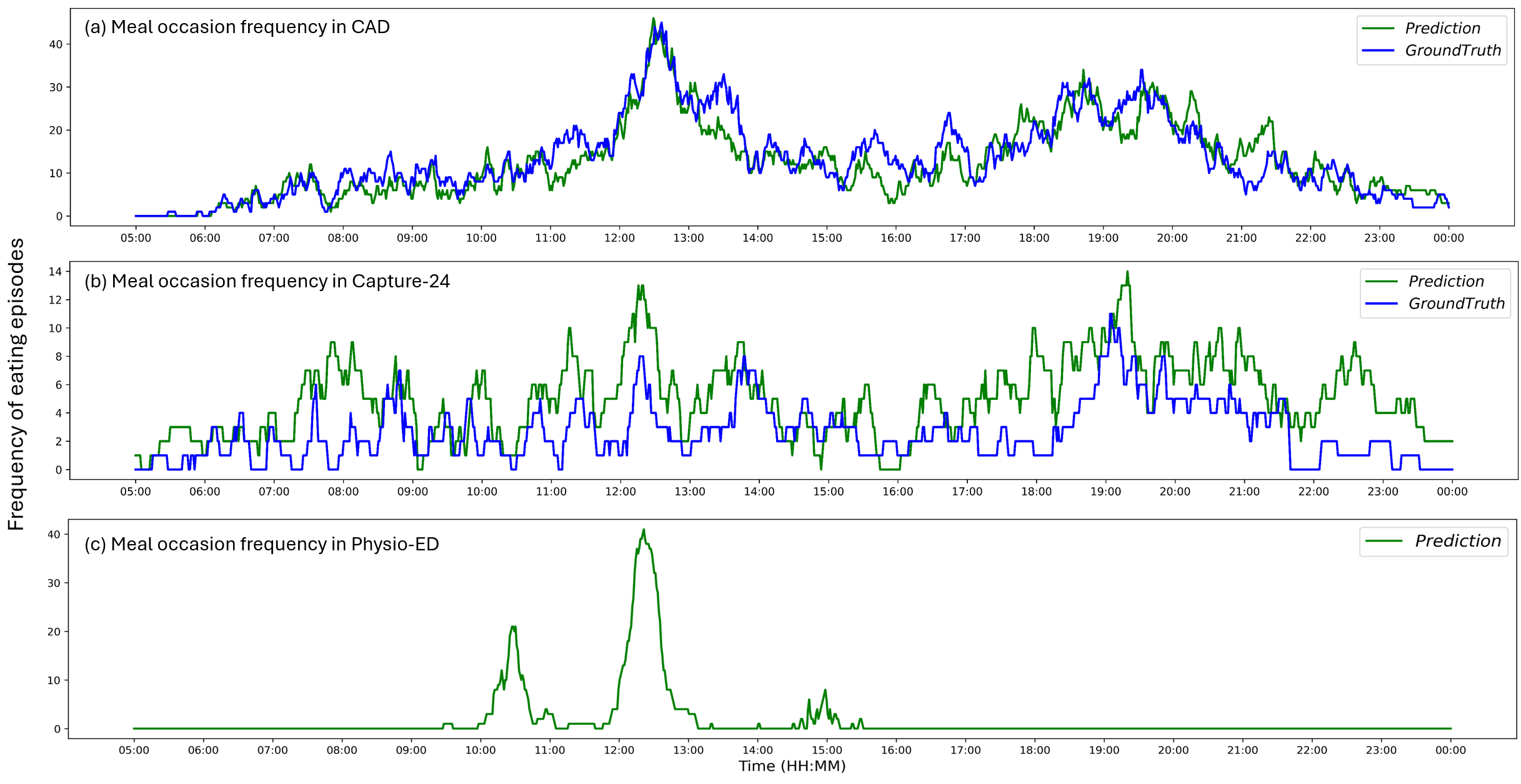}
	\vspace{-0.3cm}
	\caption{Meal occasion frequency distributions for ground truth and predicted eating episodes across three datasets. For the Physio-ED dataset, precise ground-truth timestamps are unavailable; however, as data were collected in a hospital setting, all participants followed a standardized schedule for lunch and snacks.} 
	\label{meal_frequency}
	\vspace{-0.3cm}
\end{figure*}

\subsection{External datasets} 
The CAD dataset was used as the primary external validation benchmark, as it is one of the largest and most widely used datasets for eating episode detection. Surprisingly, the performance achieved under the proposed external validation scheme substantially exceeded the state-of-the-art (SOTA) results reported using internal validation \cite{tang2023detecting}. As shown in Table~\ref{compare_litera}, the episode-level F1-score increased from 0.568 to 0.751, while the episode segmentation IoU improved from 0.38 to 0.59. These results suggest that, when sensor orientation and wearing-hand variability are appropriately addressed, bottom-up eating episode detection models can generalize across datasets more effectively than previously assumed.

The Capture-24 dataset, released in 2024, is a free-living HAR dataset and, to the best of our knowledge, has not previously been used for eating episode detection. Notably, its data collection protocol closely resembles that of the large-scale UK Biobank wearable study, which includes more than 100,000 participants. The encouraging results obtained on Capture-24 suggest that the proposed framework may be applicable to UK Biobank accelerometer data for the extraction of eating-related digital phenotypes, thereby creating new opportunities for population-level public health and epidemiological research.

Individuals with eating disorders represent an important clinical target population for automated food intake monitoring; however, this group has not previously been considered in the evaluation of eating episode detection methods. Although the Physio-ED dataset was not originally designed for eating episode detection and employs different sensing modalities, the proposed framework was nevertheless able to identify eating episodes, particularly structured meals such as lunch, as shown in Fig. \ref{meal_frequency}(c). These findings suggest that the proposed approach may provide an objective and scalable means of assessing dietary activity in the daily lives of individuals with eating disorders, addressing a significant unmet need in clinical psychiatry. Moreover, the ability to continuously monitor eating behavior in naturalistic settings could support the future development of just-in-time adaptive interventions (JITAIs) for this population.

\begin{figure*}[t]
	\centering
	\includegraphics[scale=0.27]{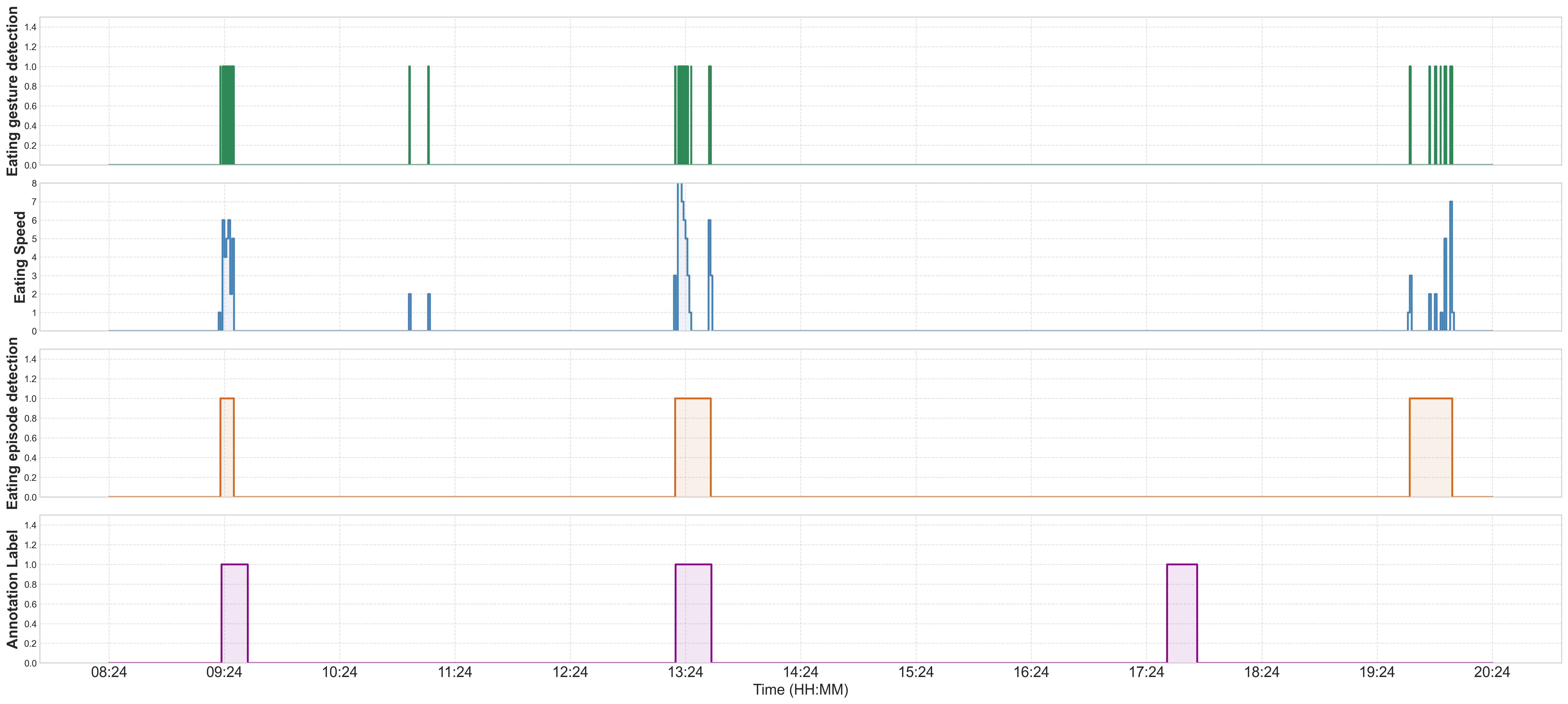}
	\vspace{-0.3cm}
	\caption{Error analysis examples from Capture-24. The first panel shows the output of eating gesture detection, the second panel shows the estimated eating speed (number of eating gestured per min) derived from the detected gestures, the third panel shows the predicted eating episodes, and the final panel shows the annotated eating episodes.} 
	\label{pre_example}
	\vspace{-0.3cm}
\end{figure*}

\subsection{Error analysis} 
Although promising results are obtained across all three datasets under the external validation scheme, several error patterns warrant further analysis. For CAD dataset, although our FP/TP rate (0.28) is substantially lower than SOTA work (1.4) \cite{tang2023detecting}, the absolute number of FNs (306) remains relatively high. An analysis of the meal occasion frequency distribution (Fig.~\ref{meal_frequency}) reveals that most FNs occur outside conventional lunch and dinner periods and during very short eating episodes, such as snacks. These episodes involve fewer and less distinctive eating gestures, making them inherently more difficult to detect reliably.

The Capture-24 dataset yields the lowest performance among the three external validation datasets, likely due to multiple contributing factors. First, the accelerometer is worn only on the dominant hand, whereas ground truth annotations are derived from a chest-mounted camera and may correspond to eating actions performed with either the dominant or non-dominant hand. As a result, eating episodes involving the non-dominant hand are inherently unobservable from the sensor data, introducing unavoidable FNs. Second, the chest-mounted camera can experience field-of-view occlusions when participants are seated at a table, potentially leading to missed or ambiguous annotations and contributing to FPs around meal times, as shown in Fig. \ref{pre_example}.

In terms of temporal segmentation accuracy, the IoU scores for detected eating episodes on both the CAD and Capture-24 datasets remain below 0.6, indicating that precise temporal alignment of eating episode boundaries remains a challenging problem. This suggests that while the framework is effective at identifying the occurrence of eating episodes, further refinement is needed to improve boundary localization.

\subsection{Limitations and future directions} 
Despite the encouraging results, several limitations should be acknowledged. First, the proposed task-driven sensor orientation inference relies on the accuracy of eating gesture detection and does not explicitly account for scenarios in which participants change sensor orientation within a single day. Addressing dynamic orientation changes remains an open challenge for future work. Second, although the proposed wearable based eating episode detection demonstrates strong technical performance for dataset from eating disorder patients (Physio-ED), its criterion validity against clinical assessments or gold standards has not yet been established. Further validation against objective or clinician-verified measures is necessary before these metrics can be adopted in clinical or nutritional research settings. Third, while the present work focuses on robust and generalizable eating episode detection in free-living environments, the more fine-grained task of generalizable eating gesture detection with reliable external validation remains unresolved. Achieving accurate gesture-level detection across heterogeneous real-world conditions continues to be a challenging but important direction for future research. Finally, the external validation datasets used in this study were collected in the US, the UK, and Belgium, and therefore primarily reflect Western-style eating behaviors (i.e., using forks and knives). Future work should extend validation to populations with diverse cultural eating practices, such as chopsticks or hands, to further evaluate the generalizability of the proposed framework, although such behaviors were partially represented in the training data.

\section{Conclusion}
This work proposes and externally validates a generalizable framework for eating episode detection using wrist-worn wearable sensors in free-living environments. By explicitly addressing sensor orientation and wearing-hand variability through a task-driven inference pipeline and adopting a two-stage gesture-to-episode detection strategy, the proposed approach demonstrates robust performance across multiple heterogeneous datasets collected under diverse protocols and from different participant populations. The framework achieves strong external validation results on a widely used benchmark dataset, extends for the first time to a general free-living activity dataset with population-scale relevance, and shows promising applicability in a clinical eating disorder cohort. Together, these findings highlight the feasibility of scalable, real-world eating episode detection and establish a foundation for future research on population-level digital phenotyping and clinically meaningful timing-related eating behavior assessment.

\section*{Acknowledgment}
The computational resources and services used in this work were provided by the VSC (Flemish Supercomputer Center), funded by the Research Foundation Flanders (FWO) and the Flemish Government – department EWI.
\bibliographystyle{IEEEtran}
\bibliography{papers}
\end{document}